\documentclass[12pt,preprint]{aastex}

\shorttitle{Fast optical variability of V648 Car}
\shortauthors{Angeloni et al.}

\begin{document}

\title{Discovery of fast, large-amplitude optical \\variability of V648 Car (=SS73-17)}

\author{R. Angeloni}
\affil{Departamento de Astronom\'ia y Astrof\'isica, Pontificia Universidad Cat\'olica de Chile, \\Vicu\~{n}a Mackenna 4860, 7820436 Macul, Santiago, Chile}
\email{rangelon@astro.puc.cl}

\author{F. Di Mille\altaffilmark{1,2}}
\affil{Sydney Institute for Astronomy, The University of Sydney\\44-70 Rosehill St., Redfern, NSW 2016, Australia}
\email{fdimille@aao.gov.au}

\author{C. E. Ferreira Lopes\altaffilmark{3}}
\affil{Universidade Federal do Rio Grande do Norte,\\ Campus Universit\'ario Lagoa Nova CEP 59078-970, Natal/RN - Brasil}
\email{carlos\_eduardo@dfte.ufrn.br}

\and
\author{N. Masetti}
\affil{INAF - Istituto di Astrofisica Spaziale e Fisica Cosmica di Bologna,\\ via Gobetti 101, 40129 Bologna, Italy}
\email{masetti@iasfbo.inaf.it}

\altaffiltext{1}{Australian Astronomical Observatory, PO Box 296, Epping, NSW 1710, Australia.}
\altaffiltext{2}{Las Campanas Observatory, Colina El Pino, Casilla 601, La Serena, Chile.}
\altaffiltext{3}{Departamento de Astronom\'ia y Astrof\'isica, Pontificia Universidad Cat\'olica de Chile.}

\begin{abstract}
We report on the discovery of large-amplitude flickering from V648 Car (= SS73-17), a poorly studied object listed amongst the very few hard X-ray emitting symbiotic stars. We performed milli-magnitude precision optical photometry with the Swope Telescope at the Las Campanas Observatory, Chile, and found that V648 Car shows large U-band variability over time scales of minutes. To our knowledge, it is amongst the largest flickering of a symbiotic star ever reported. Our finding supports the hypothesis that symbiotic WDs producing hard X-rays are predominantly powered by accretion, rather than quasi-steady nuclear burning, and have masses close to the Chandrasekhar limit. No significant periodicity is evident from the flickering light curve. The ASAS long-term V light curve suggests the presence of a tidally distorted giant accreting via Roche Lobe overflow, and a binary period of $\sim520$ days. On the basis of the outstanding physical properties of V648 Car as hinted by its fast and long-term 
optical variability, as well as 
by its nature as hard X-ray emitter, we therefore call for simultaneous follow-up  observations in different bands, ideally combined with time-resolved optical spectroscopy.
\end{abstract}


\section{Introduction}
V648 Car was a forgotten red variable star of magnitude $V \sim$ 10 until it was realized its nature as a hard X--ray emitter, i.e. when source IGR J10109-5746 was discovered at hard X--rays by the {\it INTEGRAL} satellite (Kuiper et al. 2006; Bird et al. 2010). Indeed, V648 Car lies within the error circle of this high-energy source, and  Masetti et al. (2006) first suggested that the red star could be the optical counterpart of the hard X--ray emitter, mainly on the basis of the positional correlation, subsequently confirmed by the detection of a soft X--ray counterpart with arcsecond-sized error on its position (Tueller et al. 2005).\\
The high-energy behavior of this object was explored with several X--ray studies with {\it Suzaku} (Smith et al. 2008), {\it Swift} (Kennea et al. 2009) and {\it Chandra} (Eze et al. 2010). These showed that the spectrum of the source is described with a highly absorbed thermal plasma model of temperature kT $\approx$ 10 keV with superimposed emission lines of ionized iron. Concerning the light curve, the {\it X-ray Telescope (XRT)} on board of Swift did not detect any rapid (100 to 5000 s) periodic variability, which led Kennea et al. (2009) to state that the X--rays come more likely from the boundary layer around a massive (estimated between 1.1 and 1.35 M$_\odot$) white dwarf (WD).

At these extensive studies of V648 Car in the X--ray domain it has not yet corresponded a parallel detailed investigation in the optical, if we exclude its previous inclusion in both the Sanduleak \& Stephenson's catalog of emission line objects of the southern Milky Way (where it appears as SS73-17 and is classified as a M3ep+OB star -- Sanduleak \& Stephenson 1973) and in the similar catalog by Henize (who lists SS73-17 as Hen 3-380 and describes it as having Balmer lines as well as single ionized calcium in emission -- Henize 1976). Later spectroscopic observations of 33 unclassified emission line stars from the original Sanduleak \& Stephenson's list briefly reported on V648 Car being an ``example of Mira type star [...] having H$\alpha$ and H$\beta$ in emission'' (Pereira et al. 2003).\\
The M4\,III spectral type as derived by Pereira et al. from the relative strength of the TiO spectral bands, and an apparent blue excess (see their Fig. 1), eventually brought Masetti et al. (2006) to candidate V648 Car as symbiotic star, classification further advocated by the analogy with RT Cru, a previous case of association between a symbiotic star and a newly-discovered {\it INTEGRAL} hard X--ray source (Chernyakova et al. 2005; Masetti et al. 2005; Tueller et al. 2005; Luna \& Sokoloski 2007).\\

Symbiotic stars are long-period interacting binaries composed of a hot compact star -- generally a WD -- and an evolved giant star, whose mutual interaction via accretion processes triggers the extended emission recorded from radio to X-rays. Out of $\sim200$ symbiotic stars known today (Belczy{\'n}ski et al. 2000), only 5 have been reported to emit significantly above 5 keV (V648 Car, RT Cru, T CrB, CH Cyg and MWC 560). It has been argued that the number of hard X--ray emitting symbiotic stars in the Galaxy may actually be much larger, and they have also been proposed as candidates for Ia supernovae progenitors because all the systems known to date seem to host WDs whose masses approach the Chandrasekhar limit (Eze 2011). 

In the framework of an observing campaign aimed at characterizing the WD properties in different type of accreting binary stars, we have started a systematic search for flickering (i.e., stochastic photometric variations on time-scales of a few minutes) in southern symbiotic stars (Angeloni et al. 2012, in preparation). This survey has been conceived as the complementary extension of the survey by Sokoloski et al. (2001) to the southern hemisphere and to dimmer objects. For our first ``test-run'', we have recognized V648 Car as a priority target by virtue of its strong similarity with CH Cyg and MWC560, two other hard X--ray symbiotic stars known to feed powerful jets and to show optical flickering (Leedj{\"a}rv 2004). In this letter, we present the outcome of our fast photometric monitoring of V648 Car conducted with the Swope Telescope at the Las Campanas Observatory, Chile. We first describe the observations and the data reduction/analysis process in Sect. 2. The results are presented in Sect. 3, then 
discussed in Sect. 4. Concluding remarks follow in Sect. 5.

\section{The Data}
\subsection{Observations and data reduction} \label{runs}
We observed V648 Car during the night 26/27 May 2011 with the Henrietta Swope Telescope at the Las Campanas Observatory, Chile. 

Swope is a f/7.5 Ritchey-Chr\'etien 1m reflector, currently equipped with a 2048$\times$3150, 15 $\mu$m pixel SITe CCD that provides an unvignetted field of view (FoV) of 14.8$\times$22.8 arcminutes, for an equivalent pixel scale of 0.435 arcsecond/pixel. The readout time of the CCD (full frame, unbinned configuration) is 164 s. \\
We observed the target through a Johnson $U$ filter\footnote{LC-3012 -- http://www.lco.cl/telescopes-information/irenee-du-pont/instruments/website/direct-ccd-manuals/3x3-filters-for-ccd-imaging} using a sub-array of the CCD that gives an actual FoV of 14.8$\times$7 arcminutes: this allowed us to cut down the overall read-out time to 57 sec, still counting on a reasonable number of suitable comparison stars for differential photometry (Sect. \ref{photometry}). The good guiding performance of the telescope, and the option to define a loop sequence in the CCD control software, allowed us to continuously monitor the target for about 4.5 hours with a virtually uniform cadence of 112 sec. With these instrumental setup and observing strategy, we collected at the end of the run 138 frames in which the FWHM of the star's point spread function (psf) varied between 3.2 and 6 pixels (i.e., 1.4-2.6 arcseconds).
We then performed the standard data reduction procedure using the \texttt{IRAF} \textit{ccdproc} package.

\subsection{Data analysis: differential CCD photometry} \label{photometry}
\begin{figure}
\epsscale{1}
\plotone{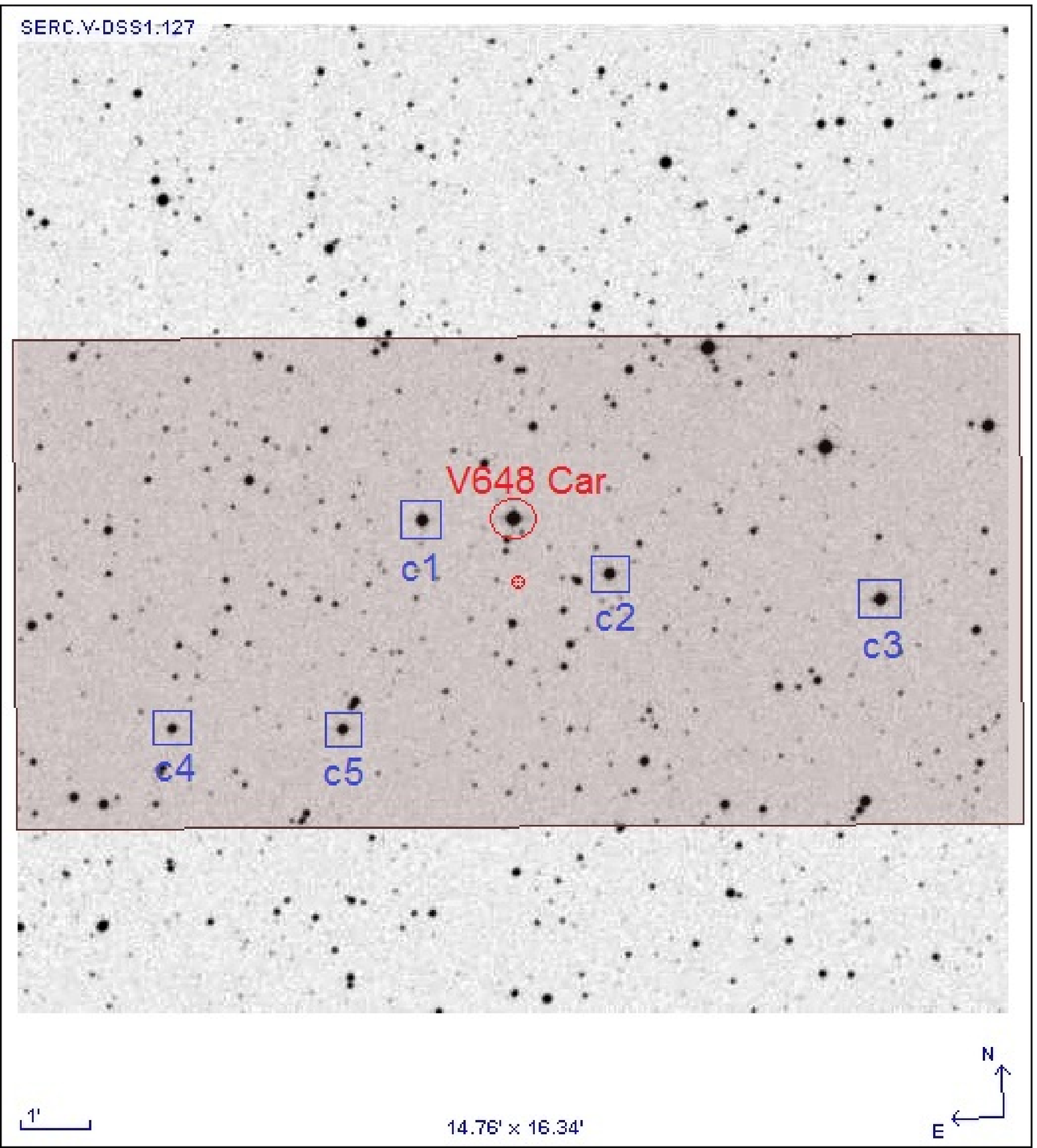}
\caption{V-band finding chart of V648 Car. The shaded area corresponds to the actual FoV after the windowing of the CCD. The red circle indicates the target, while the blue boxes mark the position of the comparison stars ($c_1$,...,$c_5$).\label{fig1}}
\end{figure}

Differential aperture photometry is a powerful yet relatively simple technique, able to deliver high-precision results even with fairly common set-up: as a matter of fact, milli-magnitude precision is nowadays routinely achieved using ground-based 1-m class telescopes and standard CCD cameras (L\'opez-Morales 2006). By simultaneously comparing the light of two or more stars in close proximity, and assuming that star psf does not vary across the detector field, one can actually deal pretty well with the error budget produced by the variations in background flux and sky brightness. For a quantitative discussion of these aspects please refer to, e.g., Milone \& Pel (2011).\\
In our case, we chose 5 comparison stars that, at a first visual inspection, qualified as good candidates for differential photometry: they are well-isolated stars, have a brightness comparable to the variable, and are located at small angular distance from the target, thus bringing a negligible atmospheric extinction difference (see Fig. \ref{fig1} \& Table \ref{table1}). 

The data analysis was performed with \texttt{VAPHOT} (Deeg \& Doyle 2001), an aperture photometry package specifically developed for dealing with precise time-series photometry of uncrowded fields. Its most important feature is the ability to work with optimized (through a preliminary psf fitting) sized apertures, calculated to extract the best S/N ratios for \textit{each} star in a CCD frame. This feature turned out to be extremely valuable in our case because of the variable seeing that we experienced along the observing night. Please refer to Deeg \& Doyle (2001) for further details about \texttt{VAPHOT} and its overall performance.  

From a first analysis of the comparison star photometry, we recovered a $U$ band atmospheric extinction coefficient $K_U=0.43\pm0.01$ in agreement with the typical value at Las Campanas and La Silla sites (Minniti et al. 1989), proving that the night was indeed photometric. The standard deviation of the light curves obtained by combining the three best comparison stars is $\lesssim$ 4 mmag. We thus assumed this value as our typical photometric error. Because of the significantly lower magnitude of the remaining $c_4$ and $c_5$ comparison stars, that only deteriorate the quality of the final photometry, we decided not to employ them further in the generation of the differential light curve of the target star.\\

\section{Results} \label{results}
\begin{figure}
\epsscale{1}
\plotone{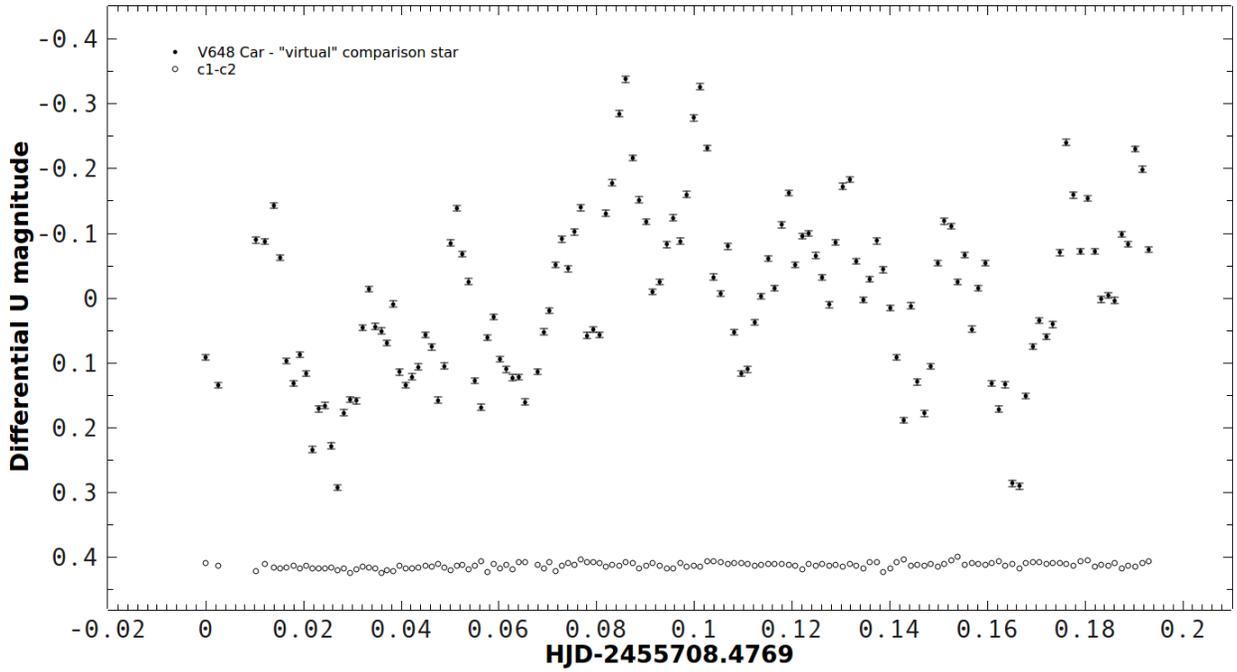}
\caption{Differential light curves of V648 Car with respect to a ``virtual star'' obtained by combining the 3 best comparison stars. The cadence of the observations is $\sim$112 s, and the maximum recorded amplitude amounts to 0.63 mag. For the sake of comparison, the \textit{$c_1$-$c_2$} differential light curve is also shown. 
\label{fig3}}
\end{figure}

In Fig. \ref{fig3} we present the differential light curve of V648 Car (black point) calculated with respect to a ``virtual star'' obtained by combining the 3 best comparisons discussed in the previous section.

During the 4.5 hours of continuous monitoring, V648 Car showed a remarkably large flickering with a recorded maximum amplitude of 0.63 mag, and with variations $>0.5$mag over a period shorter than 15 minutes. To our knowledge, this is amongst the largest flickering ever reported from a symbiotic star, comparable to the flickering amplitude of the well-studied symbiotic CH Cyg during its 1998 active state (Sokoloski \& Kenyon 2003, Contini et al. 2009).

\subsection{Period analysis}
We have performed a period analysis of our light curve with a Lomb-Scargle Periodogram (Lomb 1976; Scargle 1982): first of all, we calculated the periodicity pattern as due to our observing sampling (i.e., the spectral window) and removed the corresponding period and its first harmonic.
The frequency analysis has been limited within the range $f_0 = 1/T_{Max}$ and $f_{Max} = 100$, where $T_{Max}$ is the total time span of the observations, with a frequency step $\Delta f = 0.1/T_{Max}$. To check the significance of the several peaks populating the power spectrum, Scargle (1982) “False Alarm” criteria was used to calculate the \textit{False Alarm Probability} (FAP).
No significant peaks have been recovered in the power spectrum. Additional observations with both a higher cadence and on a longer time baseline are urgently requested to clarify whether a suspicious periodicity at $P\sim 25$ minutes might be real or is just arising from instrumental and/or detection-statistics effects.

\subsection{ASAS V-band light curve}
\begin{figure}
\epsscale{1}
\plotone{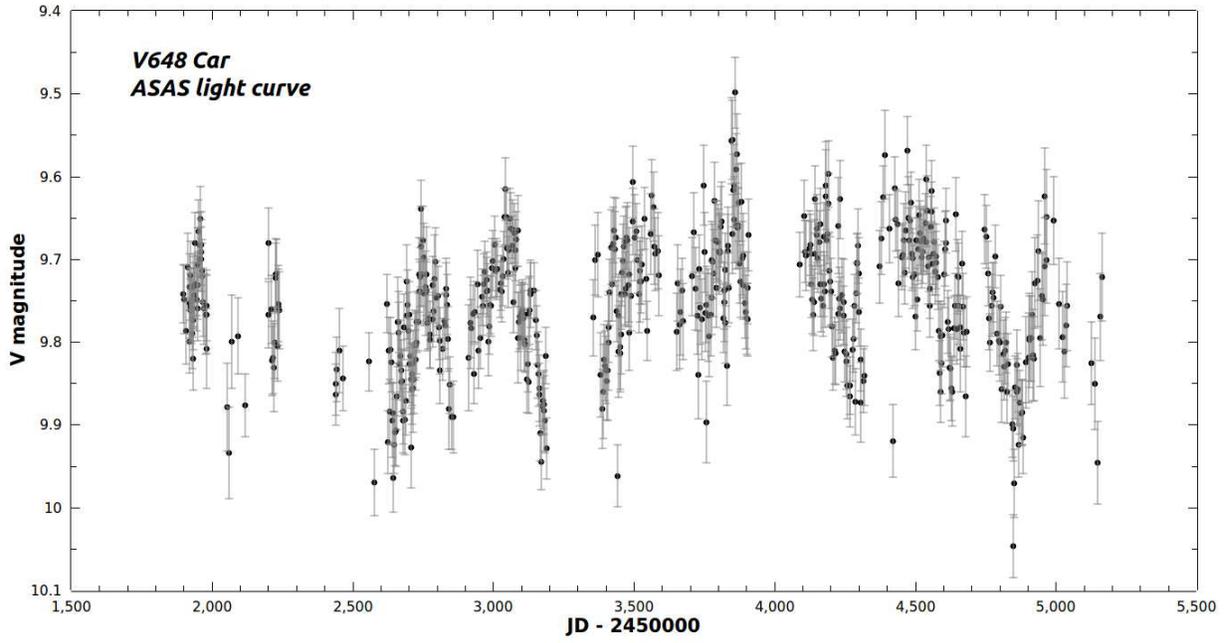}
\caption{ASAS light curve of V648 Car. The V-band amplitude amounts to 0.548 mag.\label{fig5}}
\end{figure}

In symbiotic stars, large amplitude flickering seems to emerge during active states of the system, while during quiescent states only low-level flickering (if present at all) has been recorded (e.g., Zamanov et al. 2010). To check whether V648 Car was in an active state during May 2011, we searched for long-term light curve into public photometric data services, and found a positive entry in the \textit{All Sky Automated Survey} (ASAS) Catalog, where V648 Car is reported as a unclassified variable star of period  P=577 days (Pojmanski 2003).  Unfortunately, the ASAS light curve does not extend over the time of our flickering observations, but already a quick-look at the data reveals some interesting, somehow unexpected properties. \\The ASAS V-band light curve is shown in Fig. \ref{fig5}. 
No hint of classical symbiotic activity is recognizable (as for example in the prototype Z And, Fig 3 of Skopal et al. 2012), even if a periodic modulation is evident. Most surprisingly, the light curve shape and overall amplitude ($\Delta V=0.55$)\footnote{For example, a Mira star in the General Catalog of Variable Stars is defined as to show light amplitudes from 2.5 to 11 mag in V -- http://www.sai.msu.su/gcvs/gcvs/iii/vartype.txt} are NOT typical of a Mira variable in a symbiotic star (Gromadzki et al. 2009), in striking disagreement with the Pereira et al.'s (2003) conclusion about the nature of the giant companion in V648 Car. 

A quantitative analysis of the ASAS light curve with a Lomb-Scargle Periodogram returns two highly-significant (FAP$<0.01$) periods, at $P_1=273\pm18$ days and $P_2=520\pm68$ days (the latter reminiscent of the Pojmanski's period). We start noticing that $P_2 \cong 2P_1$: while at a first sight one may be tempted to consider $P_2$ as an alias, or $P_1$ as the $P_2$ first harmonic, a careful visual inspection of the data brings us to suggest a tentative scenario (exemplified by the light curve trend between $\sim$JD2452600 and $\sim$JD2453200) in which $P_2=520$ days would represent the true orbital period of the system, while $P_1=273$ days would mirror the ellipsoidal distortion of the giant. While the former period would basically be recoverable thanks to the reflection effect of the cool giant's hemisphere that is facing the WD (effect that is more evident at shorter wavelengths), the ellipsoidal variation due to the giant would be more easily detected at longer wavelengths. 
If this scenario is correct, the giant is then at the inferior conjunction at $\sim$JD2452649 (phase $\phi$=0); the two maxima at $\sim$JD2452749 and $\sim$JD2453043 mark instead the quadratures (phases $\phi$=0.25 and $\phi$=0.75, respectively), when the ellipsoidal effect is expected to be maximal.
In the V band these two effects compensate each other, and the resultant light curve thus displays a smoothed profile, able to explain the reduced depth of the secondary minimum occurring at $\phi$=0.5 ($\sim$JD2452874), when the ellipsoidal effect is minimal, but the reflection effect is maximal. For a review of the various types of variability seen in symbiotic stars, see e.g. Munari (2012) and Skopal (2008).\\ 
First-order ephemeris for V648 Car would thus take the form:
\begin{equation}
HJD = 2452649(\pm1) + E \times 520(\pm68) 
\end{equation}

Even if tentative, our scenario can be easily falsified with additional light curves in a redder filter (suitable would be the I filter, ideal some near-infrared filters such as the JHK ones) and/or a radial velocity curve: in the former case, we would expect to detect $P_1$, while in the latter case only $P_2$ should be recovered. It is worth explicitly mentioning that, even if the reflection effect is stronger in the ultraviolet, it would be almost impossible to recover any orbital period via U band photometric monitoring: the amplitude of the U band flickering displayed by V648 Car is at such a level that one point per run cannot be taken as representative of the daily magnitude of the system.

\section{Discussion} \label{discussion}
Amongst an overall population of about 200 symbiotic stars, less than 5 members are reported as large flickering symbiotics (Gromadzki et al. 2006): the northern survey by Sokoloski et al. (2001) seemed to confirm a general lack of flickering activity in the majority of the monitored systems, down to a few mmag level. Even if a quantitative physical model does not exist yet, different interpretations have been discussed in the literature to explain the physical and spatial origin of flickering. With no aim of completeness, here we just mention two vaguely opposite scenarios that
set the so-called propeller-accretor model (Mikolajewski et al. 1996), for which the flickering results from the interaction of the magnetic WD magnetosphere with the giant's wind, against the Sokoloski \& Kenyon (2003) hypothesis that, at least in the case of the large-amplitude flickerer CH Cyg, this phenomenon actually originates in the inner parts of a (probably thermally unstable) accretion disk. \\
If our interpretation of the ASAS light curve in terms of ellipsoidal variability is correct, then we expect that the V648 Car orbital plane has a moderate-to-high inclination and might actually be an eclipsing system. By accepting the paradigm of flickering as originating at the inner edge of the accretion disk (i.e., the boundary layer), we could thus expect it to disappear when the giant is at the inferior conjunction ($\phi$=0) and the WD and/or the accretion disk then eclipsed. Incidentally, the Swope observing run was performed around $\phi$=0, but the accuracy of our ephemeris combined with just one set of observations does not allow to establish whether V648 Car is an eclipsing symbiotic binary. It would however be interesting to check through repeated monitoring if flickering activity is still present (and at which level) when the giant is at the next inferior conjunction: on the basis of Eq. 1, this will happen around January 2013. 

The lack of coherent rapid periodic variability in both X--rays and optical light curves of symbiotic stars has been usually reported as an evidence that in these systems the white dwarf is non-magnetic (Mukai et al. 2008). The unique exception seems to be Z And, which displays a persistent oscillation at $P\sim28$ min (with typical amplitude of 0.02 mag in U) naturally explained as arising from the rotation of a highly-magnetic WD (Sokoloski \& Bildsten 1999). In a few other symbiotic flickers, not-strictly coherent periodicities have been reported in the past (e.g., $P\sim22$ min in MWC560 -- Dobrzycka et al. 1996), but there is not a general consensus yet about their real existence, nor in turn about the magnetic character of the accreting WD (Zamanov \& Raikova 1999).
Similarly, we could not recover any significant periodicity in our V648 Car data, and the observed variability seems indeed of stochastic nature. Further observations are needed to conclude about the intrinsic nature of a suspected periodicity at $P\sim 25$ minutes, for which an instrumental and/or detection-statistics origin cannot be discarded at the moment.


In the ASAS light curve we recognized an ellipsoidal modulation interpreted as the signature of a tidally distorted giant companion: in this scenario, the accretion would occur via the more effective Roche-Lobe overflow (RLO), and not via wind, like in the majority of symbiotic stars, therefore implying a higher mass accretion rate. We also interpret the long-term modulation of $\sim520$ days visible in the ASAS data as the binary orbital period.

Considering its nature of hard X-ray emitter, and the remarkable properties of the flickering that we discovered, V648 Car bears outstanding similarities with other well-known symbiotic stars (e.g., T CrB, CH Cyg, MWC560) known also to host massive WDs and to have experienced outburst activity in the recent past. It appears thus reasonable to extend theses similarities by suggesting that in the future V648 Car will likely experience an outburst typical of a symbiotic nova (Mikolajewska 2010), while the inferred high-mass of its WD brings additional weight to the candidature of symbiotic stars as progenitors of type Ia supernovae (Chen et al. 2011, Chiotellis et al. 2012).

\section{Concluding remarks} \label{discussion}
We have discovered fast, large-amplitude flickering of V648 car, a poorly studied object listed amongst the very few hard X-ray emitting symbiotic stars. With an overall U-band amplitude of $\sim0.6$ mag over time scales of minutes, it is one of the largest flickering ever reported from a symbiotic star. 
This finding is a further confirmation that symbiotic WDs producing hard X-rays are predominantly powered by accretion (rather than quasi-steady nuclear burning on the surface of the WD), and therefore that the insights about the masses of these WDs based on this interpretation are fundamentally correct.
The ASAS light curve suggests the presence of a tidally distorted giant accreting via RLO, and a binary period of $\sim520$ days. 
The outstanding character of V648 Car, as reconstructed from the puzzle of its fast and long-term optical variability and by its hard X-ray emitting properties, clearly demands a coordinated follow-up: e.g., simultaneous X/UV/optical observations in different bands, ideally combined with time-resolved optical spectroscopy.

\acknowledgments
The authors would like to thank an anonymous referee for constructive criticism that improved the paper. Support for R.A. is provided by Proyecto Fondecyt \#3100029. F.D.M. acknowledges the support of a Magellan Fellowship from Astronomy Australia Limited, and administered by the Australian Astronomical Observatory.  C.E.F.L acknowledges support from the CAPES Foundation, Ministry of Education of Brasil, Brasilia DF 70040-020, Brasil (Processo \#5266/11-4). This paper includes data gathered with the 1m Swope Telescope at the Las Campanas Observatory, Chile. It also made use of ASAS data.

{\it Facilities:} \facility{Facility: Swope}.

\clearpage

\begin{table}
\begin{center}
\caption{Comparison stars used for the differential photometry (see also Fig. \ref{fig1})\label{table1}.}
\begin{tabular}{ccccccccccccccccccccc}
\tableline\tableline
Star & RA (J2000) & DEC (J2000) & V mag & B-V & d\tablenotemark{a}  \\
\tableline
c1 &10 11 12.92 & -57 48 16.14 &  10.96 & 0.29 &1.3 \\
c2 &10 10 52.69 & -57 49 02.07 &  10.81 & 0.29 &1.6 \\
c3 &10 10 23.64 & -57 49 24.54 &  10.56 & 0.64 &5.4 \\
c4 &10 11 39.81 & -57 51 12.48 &  12.20 & 0.3 &5.7\\
c5 &10 11 20.97 & -57 51 17.39 &  12.63 & -0.17 &3.9 \\
\tableline
\end{tabular}
\tablenotetext{a}{Distance from the target, in arcminutes.}

\end{center}
\end{table}

\end{document}